\documentclass[aps,a4paper, reprint, twoside]{revtex4-1}
\pdfoutput=1

\usepackage[english]{babel}
\usepackage[x11names]{xcolor}
\usepackage{float}
\usepackage[pdftex,hidelinks,unicode=true, pdftitle={Article}, pdfauthor={Author}]{hyperref}
\hypersetup{colorlinks=true, citecolor=black, linkcolor=black, urlcolor=blue} 
\usepackage{mathtools}
\usepackage{latexsym,amsthm, amssymb}
\usepackage[pdftex]{graphicx}
\usepackage{lipsum}
\usepackage[normalem]{ulem}

\DeclarePairedDelimiter\ton{(}{)}
\DeclarePairedDelimiter\qua{[}{]}
\DeclarePairedDelimiter\gra{\{}{\}}

\DeclarePairedDelimiter\av{\langle}{\rangle}

\newcommand{\km}{k_{\mathrm{min}}}
\newcommand{\kM}{k_{\mathrm{max}}}

\newcommand{\be}{\begin{equation}}
\newcommand{\ee}{\end{equation}}
\newcommand{\bea}{\begin{eqnarray}}
\newcommand{\eea}{\end{eqnarray}}
\newcommand{\change}[1]{{\color{black}{#1}}}

\begin{document}

\title{Degree-Ordered-Percolation on uncorrelated networks}
\author{Annalisa Caligiuri}
\affiliation{Dipartimento di Fisica Universit\`a ``Sapienza”, P.le A. Moro, 2, I-00185 Rome, Italy.}

\author{Claudio Castellano}
\affiliation{Istituto dei Sistemi Complessi (ISC-CNR), via dei Taurini 19, I-00185 Rome, Italy}
\email[email address: ]{claudio.castellano@roma1.infn.it}

\date{\today} 
\begin{abstract}
  We analyze the properties of Degree-Ordered Percolation
  (DOP), a model in which the nodes of a network
  are occupied in degree-descending order.
  This rule is the opposite of the much studied degree-ascending protocol,
  used to investigate resilience of networks under intentional attack,
  and has received limited attention so far.
  The interest in DOP is also motivated by its connection with the
  Susceptible-Infected-Susceptible (SIS) model for epidemic spreading,
  since a variation of DOP is related to the
  vanishing of the SIS transition for random
  power-law degree-distributed networks $P(k) \sim k^{-\gamma}$.
  By using the generating function formalism, we investigate the behavior
  of the DOP model on networks with generic value of $\gamma$ and
  we validate the analytical results by means of numerical simulations.
  We find that the percolation threshold vanishes in the limit of large
  networks for $\gamma \le 3$, while it is finite for $\gamma>3$, although its
  value for $\gamma$ between 3 and 4 is exceedingly small
  and preasymptotic effects are huge.
  We also derive the critical properties of the DOP transition, in particular
  how the exponents depend on the heterogeneity of the
  network, determining that DOP does not belong to the universality
  class of random percolation for $\gamma \le 3$.
\end{abstract}

\maketitle

\section{Introduction}

The investigation of percolative properties of complex networks
has attracted a huge interest over the past
20 years~\cite{Newmanbook,Dorogovtsev2008,Lee2018}.
Many highly nontrivial phenomena occur in this
context, including continuous, discontinuous and hybrid transitions.
The first pioneering investigations pointed out the strong effect 
of the degree distribution, making heterogeneous structures very
resilient with respect to random failures but extremely fragile under
intentional attacks targeted at the most connected
elements~\cite{Cohen2000,Callaway2000,Albert2000,Cohen2001}.
A general model to investigate the effect on the percolation transition
of degree-dependent protocols for removing network elements (nodes)
was introduced by Gallos {\em et al.}~\cite{Gallos2005}.
In that model, at each time step the probability that a node of degree
$k_i$ is removed is proportional to $k_i^{-\alpha}$. The cases $\alpha=0$
and $\alpha \to +\infty$ correspond to standard random percolation and to
intentional attack, respectively. The case $\alpha \to -\infty$ corresponds
to a percolation process where nodes are added in degree-descending
order (or alternatively, removed in degree-ascending order).
This process was called Degree-Ordered-Percolation (DOP) by
Lee {\em et al.}~\cite{Lee2013}, who considered it in the context
of the debate about the asymptotic properties of the
Susceptible-Infected-Susceptible (SIS) model for epidemics on
networks with degree distribution
$P(k) \sim k^{-\gamma}$~\cite{PastorSatorras2015}.
They argued that if the DOP threshold vanishes in the large-network limit
for $\gamma>3$
then this would imply that also the SIS threshold should vanish in
the same limit.
Hence the vanishing of the DOP threshold for $\gamma>3$ would have reconciled
theoretical arguments suggesting a finite SIS threshold~\cite{Goltsev2012}
with numerical results showing it to be
vanishing~\cite{Boguna2013}.
Lee et al. studied DOP numerically and found a finite DOP
threshold for $\gamma>4$ and less conclusive evidence for $3<\gamma<4$.
These results indicated that DOP is not at the origin of the vanishing
of the SIS threshold observed numerically for $\gamma>3$.
See Ref.~\cite{PastorSatorras2015} for more details.

Despite this lack of a direct connection with the SIS transition,
DOP is a simple
and interesting model whose properties have not been, to the best of our
knowledge, fully understood. The only analytical investigation
was performed by Lee {\em et al.} about DOP on some peculiar
hierarchical scale-free flower networks~\cite{Lee2014}.
It is natural to wonder what is the expression for the DOP threshold
as a function of $\gamma$
and whether the critical exponents are different from those of standard
random percolation.
Moreover, very recent
work~\cite{Castellano2020} has shown that SIS dynamics is actually
connected to a long-range type of process, Cumulative Merging Percolation,
of which DOP constitutes the nontrivial short-range limit.
For these reasons in this paper we reconsider DOP on power-law
degree-distributed networks and by means of analytical and
numerical results we fully clarify its phenomenology.

\section{The model}
The Degree-Ordered-Percolation model is defined as follows.
We consider a generic network and start removing nodes in degree-ascending
order, i.e. we start from the nodes with smallest degree $\km$ and once
they are all removed we start removing nodes with degree $\km+1$ and so
on. Nodes with the same degree are removed in random
order. The process can also be seen as starting from a network where all
nodes have been removed and iteratively putting them back in degree-descending
order.
It is clear that this process is exactly the opposite of the much
studied percolation process under intentional attack,
investigating network robustness when nodes are removed starting from
the most connected ones~\cite{Albert2000,Cohen2001}.

Let us define $p$ as the ratio between the number of nodes added
(or not removed) and the total number $N$ of
nodes in the original network. 
The quantity $p$ is the control parameter in our system.
For $p=1$ the topology is the original one, that we assume to be
connected (i.e., all nodes belong to the giant connected component).
For $p=0$ all nodes have been removed and the relative size of the largest
connected component is null.
An intermediate value $p=p_c$ marks the birth of an extensive giant
connected component.
Our goal is to determine how this quantity and the related critical
behavior depend on the network properties.
In particular, we study this percolation process for
power-law distributed uncorrelated networks
where the normalized degree distribution is, for a finite size network
\be
P(k) = \frac{\gamma-1}{\km^{1-\gamma}-\kM^{1-\gamma}} k^{-\gamma}.
\label{pk}
\ee
As minimum degree we take $\km=3$, while the maximum degree
is set equal to $\kM=N^{1/2}$ for $2<\gamma \le 3$ and $\kM=N^{1/(\gamma-1)}$ for
$\gamma>3$.

\section{The percolation threshold $p_c$}
In order to determine the percolation threshold we follow the argument
of Ref.~\cite{Gallos2005}. We apply the general Molloy-Reed
criterion~\cite{Molloy95},
stating that a giant component exists provided the network branching
factor is larger than 1
\be
\sum_k \frac{k^2-k}{\av{k}} P_p(k) > 1,
\label{MR}
\ee
where $P_p(k)$ is the degree distribution of the network for a given value of
the control parameter $p$. For DOP this distribution is simply given by
\be
P_p(k) = \Theta(k-k_p) P(k),
\ee
where $k_p$ is the minimum degree of nodes still left in the network
and $\Theta(x)$ is the Heaviside step function.
\change{The inequality in Eq.~\eqref{MR} becomes an equality} for a critical value
$k_c$ of the degree $k_p$, which determines the onset of a
giant connected component in the system, i.e., the \change{percolation} transition.
Once $k_c$ is known, the \change{percolation} threshold $p_c$ is determined by
the condition
\be
\sum_k  \Theta(k-k_c) P(k) = p_c.
\label{pc}
\ee
These expressions correspond to the limit $\alpha \to -\infty$ of the
general treatment presented in Ref.~\cite{Gallos2005}.

\subsection{$\gamma>3$}

\begin{figure}
\includegraphics[width=0.49\textwidth]{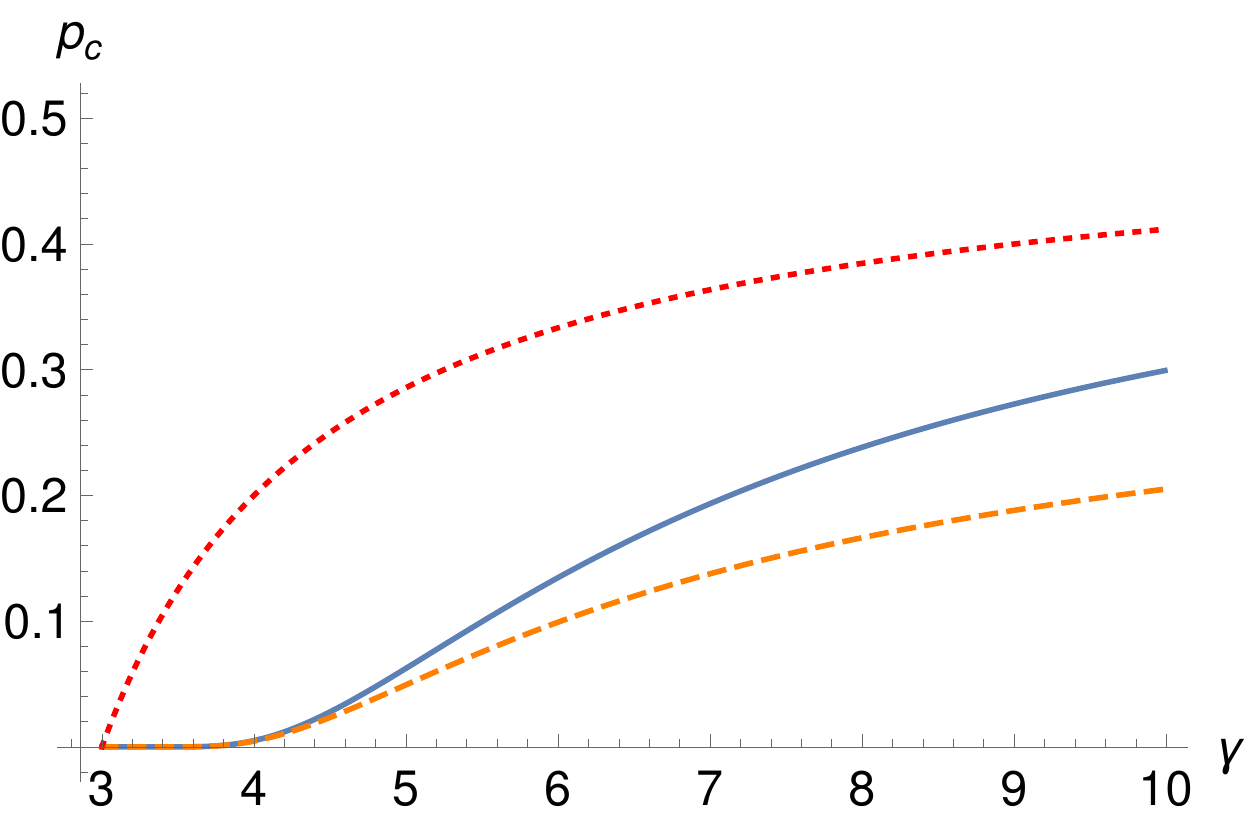}
\caption{Plot of $p_c$ as a function of $\gamma>3$. The blue solid line
  represents the exact solution, the orange dashed line represents 
  the approximate one, Eq.~\eqref{pc3}. The dotted red line is the
threshold for standard percolation.}
\label{pcvsgamma}
\end{figure}

Let us first consider the case $\gamma>3$ in the infinite size
limit $N \to \infty$.
Taking the continuous degree limit,
from Eq.~(\ref{MR}) the threshold condition is, 
\be
\int_{\km}^{\infty} dk \frac{k^2-k}{\av{k}} \Theta(k-k_c) P(k) = 1,
\label{MR2}
\ee
which, reminding that $\av{k} = \frac{\gamma-1}{\gamma-2} \km$, yields
\be
\frac{\gamma-2}{\gamma-3} \km^{\gamma-2} k_c^{2-\gamma}\ton*{k_c-
  \frac{\gamma-3}{\gamma-2}} = 1.
\label{kc}
\ee
Hence, for any $\gamma>3$, an extensive giant component appears as soon
as nodes down to the finite degree $k_c$ are added.
Solving this equation numerically, inserting the result into
Eq.~\eqref{pc}, we obtain (again assuming the continuous degree limit)
the value of the finite \change{percolation} threshold
\be
p_c = \ton*{\frac{\km}{k_c}}^{\gamma-1},
\label{pc2}
\ee
which is displayed (as a solid line) in Fig.~\ref{pcvsgamma}.
Note that in the range between $\gamma=3$ and $\gamma=4$ the value
of $p_c$ is exceedingly small but it is not equal to 0. This result
provides solid evidence about an issue that was not completely
clarified by numerical simulations in Ref.~\cite{Lee2013}.

From Eq.~\eqref{kc} one can obtain an explicit approximate expression
for $k_c$ by neglecting $(\gamma-3)/(\gamma-2)$ with respect to $k_c$.
Inserting such expression into Eq.~\eqref{pc2} yields 
\be
p_c = \ton*{\frac{\gamma-3}{\gamma-2}\frac{1}{\km}}^{\frac{\gamma-1}{\gamma-3}}.
\label{pc3}
\ee
As expected, the approximation works well for relatively small values
of $\gamma$, see Fig.~\ref{pcvsgamma}.

It is also possible to derive the finite size corrections to the expression
of $p_c$.
As shown in Appendix~\ref{fss},
for a network of finite size $N$ the effective threshold $p_c(N)$ is
\be
p_c(N)-p_c \propto \kM^{3-\gamma} \propto N^{-\frac{\gamma-3}{\gamma-1}}.
\ee

The critical exponent $\nu$, defined by
$p_c(N)-p_c \sim N^{-1/\nu}$ is then $\nu=(\gamma-1)/(\gamma-3)$.
This estimate is valid only up to $\gamma=4$.
Above that $\gamma$ value the correction to $p_c$ becomes subleading with
respect to the $N^{-1/3}$ correction due to critical fluctuations,
present also in homogeneous systems~\cite{Dorogovtsev2008}.

\subsection{$2<\gamma \le 3$}
In this case, the sum appearing in Eq.~\eqref{MR} diverges in the limit
of infinite size if $k_p$ remains finite. This indicates that $k_c$
should be diverging for large $N$. Let's analyze this in detail.
We still consider the continuous limit \change{as in} Eq.~\eqref{MR2}
but for a finite
network we perform the integral only up to $\kM$, obtaining
\be
\frac{\gamma-2}{\gamma-3} \km^{\gamma-2} \ton*{k_c^{3-\gamma}-\kM^{3-\gamma}}= 1,
\ee
where we have neglected the term proportional to $k$ in the sum, as
it remains finite in the infinite size limit. This implies that
\be
k_c = \kM \ton*{1-\frac{\km^{2-\gamma}}{\kM^{3-\gamma}}\frac{3-\gamma}{\gamma-2}}^{1/(3-\gamma)}.
\ee
To evaluate the percolation threshold one has to take into account the
finite size also in Eq.~\eqref{pc}, obtaining
\be
p_c(N) = \ton*{\frac{\km}{k_c}}^{\gamma-1} \qua*{1-\ton*{\frac{\kM}{k_c}}^{1-\gamma}}.
\label{pcf}
\ee
Inserting the expression for $k_c$ into Eq.~\eqref{pcf}, after some
algebra we obtain
\be
p_c(N) = \frac{\gamma-1}{\gamma-2} \km \kM^{-2} = \frac{\gamma-1}{\gamma-2} \km
N^{-1}.
\label{pc4}
\ee
Thus we find that the threshold vanishes in the infinite size limit and
$\nu=1$.
The inverse proportionality between $p_c(N)$ and $N$ in Eq.~\eqref{pc4} leads
to the surprising conclusion that the incipient giant component
at $p_c(N)$ is composed by a {\em finite and very small} number of nodes.
For example,
for $\km=3$ this number is only $9$ for $\gamma=2.5$ and tends to $6$
for $\gamma \to 3$.

\change{
In Appendix~\ref{fss} we present the calculation of the asymptotic value
of the threshold and of its finite size corrections for the case $\gamma=3$.
We find that $p_c(N)$ decays to 0 with $N$ with exponent $1/\nu=1$, as in
Eq.~\eqref{pc4}, but with a prefactor $k_{min}^2(e^{2/k_{min}}-1)$ which is different
from the limit $2 k_{min}$ of Eq.~\eqref{pc4} for $\gamma \to 3^-$.
}

Summarizing, we find that the exponent $1/\nu$ governing how the effective
threshold $p_c(N)$ approaches its infinite size limit is
\be
\frac{1}{\nu} =
\begin{cases}
  1~~~~~~~~~~~  \text{for } 2<\gamma\, \change{ \le}\, 3\\
  \frac{\gamma-3}{\gamma-1} ~~~~~~~\, \text{for } 3<\gamma<4\\
  \frac{1}{3}~~~~~~~~~~~  \text{for } \gamma>4.
\end{cases}
\ee

\section{Critical exponents of the percolation transition}

\subsection{The exponent $\beta$}
We want to determine how $|G(p)|$, the relative size of the giant component,
grows in the vicinity of the percolation threshold
\be
|G(p)| \sim \Delta^{\beta},
\ee
where $\Delta=p-p_c$ is the distance from the critical point.
We make use of the generating function formalism, a standard tool for
percolation problems in networks~\cite{Newmanbook}.
Indicating with $u$ the probability that a node is not connected to the
giant component $G$ through one of its neighbors, the generating functions
are defined as
\be
f_0(u) = \sum_k P_p(k) u^k
\ee
and
\be
f_1(u) = \sum_k \frac{k P_p(k)}{\av{k}} u^{k-1}.
\ee
Given these definitions the size of the giant component is~\cite{Newmanbook}
\be
|G(p)| = f_0(1)-f_0(u)
\label{G}
\ee
where the value of $u$ is the solution of
\be
u = 1-f_1(1)+f_1(u).
\label{u}
\ee
Below the threshold $u=1$, while above it $u<1$.
Since we are interested in the vicinity of the critical point
we set $u=1-\epsilon$ and expand for small $\epsilon$.
By considering the continuous degree limit, from Eq.~\eqref{G} we find
(see Appendix~\ref{appbeta})
that for any $\gamma>2$ to leading order
\be
|G(p)| \simeq \frac{\gamma-1}{\gamma-2} \km^{\gamma-1} k_p^{2-\gamma} \epsilon,
\ee
where the quantity $k_p$ is related to $p$ by
\be
k_p = \km p^{1/(1-\gamma)}.
\ee
For $\gamma>3$ the quantity $k_p$ goes to the finite value $k_c$
at the transition so that
the critical behavior is determined only by the dependence of
$\epsilon$ on $\Delta$.

As shown in Appendix~\ref{appbeta},
$\epsilon \sim \Delta$ for $\gamma>4$,
while $\epsilon \sim \Delta^{1/(\gamma-3)}$ for $3<\gamma<4$.
For $2<\gamma \, \change{\le} \, 3$ instead, since $p_c=0$,  $k_p$ diverges as
$\Delta^{1/(1-\gamma)}$ close to the transition, while
$\epsilon \sim k_p^{-1}$ (see Appendix~\ref{appbeta}),
hence overall $|G(p)| \sim k_p^{1-\gamma} \sim \Delta$.
In summary, we find that the relative size of the giant component
close to the critical transition scales with an exponent
\be
\beta =
\begin{cases}
  1~~~~~~~~~~~  \text{for } 2<\gamma \, \change{ \le } \,3\\
  \frac{1}{\gamma-3} ~~~~~~~\, \text{for } 3<\gamma<4\\
  1~~~~~~~~~~~  \text{for } \gamma>4.
\end{cases}
\ee

\subsection{The exponent $\tau$}

At the percolation critical point, the probability $n_s(p)$ that a finite
cluster has size $s$ decays as
\be
n_s(p) \sim s^{-\tau}.
\ee
To determine this exponent, we consider the associated probability
that a randomly chosen node belongs to a cluster of size $s$, 
$p_s = s n_s$.
The function that generates this distribution is~\cite{Callaway2000}
\be
h_0(x) = \sum_s p_s x^s,
\ee
while $h_1(x)$ is the generating function associated to the
probability for a node to be connected to a finite cluster of size $s$
through one of its neighbors.
These two generating functions are related to the generating
functions $f_0(x)$ and $f_1(x)$ as follows~\cite{Callaway2000}:
\bea
\label{h0}
h_0(x)&=&1-f_0(1)+xf_0[h_1(x)]\\
h_1(x)&=&1-f_1(1)+xf_1[h_1(x)].
\label{h1}
\eea
\change{
In Appendix~\ref{C} we determine the behavior of the generating
functions for $x=1-\epsilon$ with $\epsilon \to 0$.
In particular, we find, defining $g_0 = 1 -h_0(1-\epsilon)$, that
$g_0 \sim \epsilon^{1/(\gamma-2)}$ for $3<\gamma<4$ and
$g_0 \sim \epsilon^{1/2}$ for $\gamma>4$, while $h_0$ and $h_1$
are trivial for $2<\gamma \le 3$.
Using Tauberian theorems~\cite{Cohen2002} we have that, if
$g_0(\epsilon) \sim \epsilon^y$, then $\tau=y+2$, leading to
\be
\tau = \left\{
\begin{array}{lll}
\frac{2 \gamma-3}{\gamma-2} & & 3 < \gamma < 4 \\
\frac{5}{2} & & 4 < \gamma.
\end{array}
\right .
\label{tau}
\ee
}

\section{Numerical results}
We check the results of the analytical approach by performing numerical
simulations of the DOP percolation process on networks built using the
uncorrelated configuration model~\cite{Catanzaro2005}.
To determine the value $p_c$ of the percolation threshold
for given $\gamma$ and $N$,
we generate many realizations of the network
and perform many realizations of the DOP process (with different random
orderings of nodes having the same degree) on each of them.
The threshold is determined from the position of the peak of the susceptibility
\begin{equation}
\chi= \frac{\sum_{s}s^2n_s}{\sum_{s'}s'n_{s'}},
\end{equation}
i.e. the mean size of the finite clusters.
The peak height $\chi_{max}$ is expected to grow with the system size
as $N^{1-2\beta/\nu}$, due to the hyperscaling relation $2 \beta+\gamma=\nu$.

\subsection*{$\gamma>4$}
To verify the validity of analytical results in this range we consider
$\gamma=4.5$.
For this value, the threshold predicted by the continuous theory
[Eqs.~\eqref{kc} and~\eqref{pc2}]
is $p_c \approx 0.0234$. However, the corresponding
value of $k_c$ is smaller than 9. With such a small range of $k$ values
taking the continuous degree limit is not appropriate. We then solve
numerically Eqs.~\eqref{MR} and~\eqref{pc},
using discrete sums and in this way we find
$p_c \approx 0.0444$.

\begin{figure}
  \includegraphics[width=0.49\textwidth]{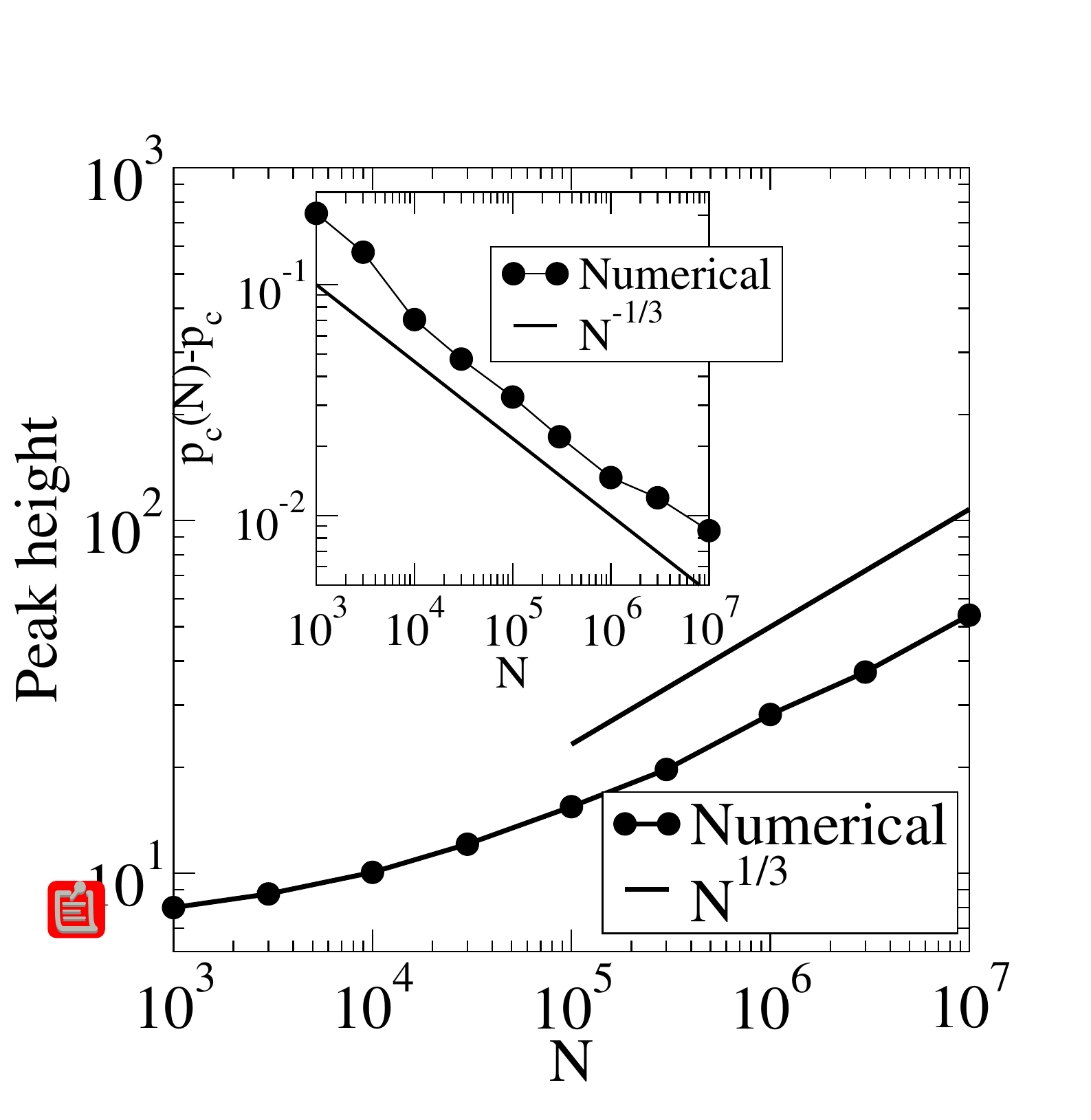}
  \caption{Results for $\gamma=4.5$. Main: Susceptibility peak height as a function of the system size
    $N$. Numerical results are compared with the theoretical prediction
    $N^{1-2\beta/\nu}$ which gives an exponent $1/3$ in this range of
    $\gamma$ values.
    Inset: Difference between the numerical effective threshold $p_c(N)$ and the
    expected value $p_c$ for infinite size, as a function of $N$.
    The straight solid line is the theoretical prediction $N^{-1/3}$.}
  \label{fss4.5}
\end{figure}

\begin{figure}
  \includegraphics[width=0.49\textwidth]{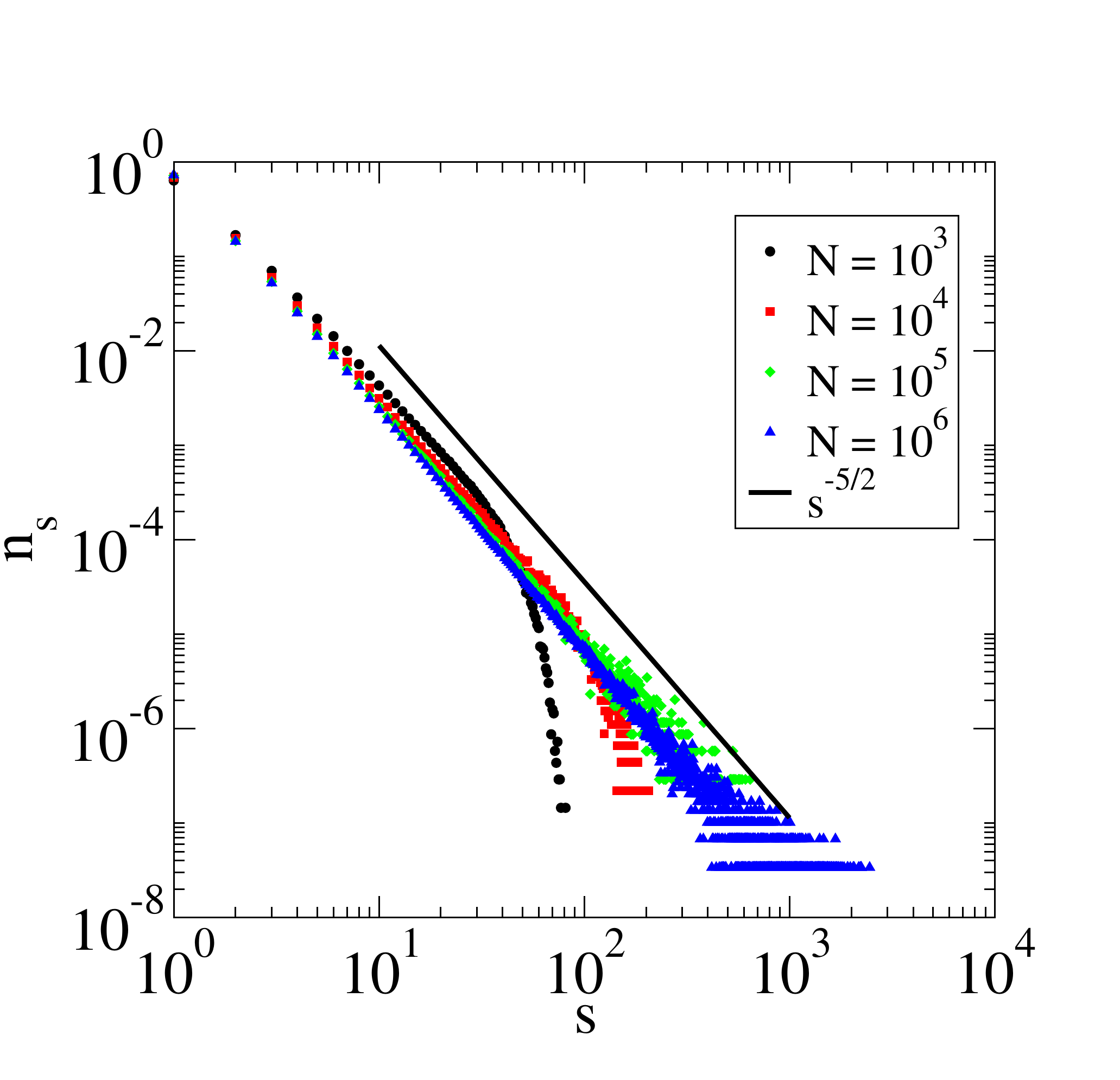}
  \caption{Cluster size distribution $n_s$ at the percolation
    threshold $p_c(N)$ for $\gamma=4.5$ and various system sizes $N$.
    The solid straight line
      represents the decay predicted analytically, Eq.~\eqref{tau}.}
  \label{ns4.5}
\end{figure}

The scaling with the system size of the peak position (effective threshold)
and peak height are displayed in Fig.~\ref{fss4.5}. The agreement between
theoretical predictions and numerical results is reasonable but not
perfect, presumably because of the discreteness of degree values mentioned
above.
The distribution of cluster sizes $n_s$ at the critical point
obeys instead very well the expected behavior (see Fig.~\ref{ns4.5}).

\begin{figure}
  \includegraphics[width=0.49\textwidth]{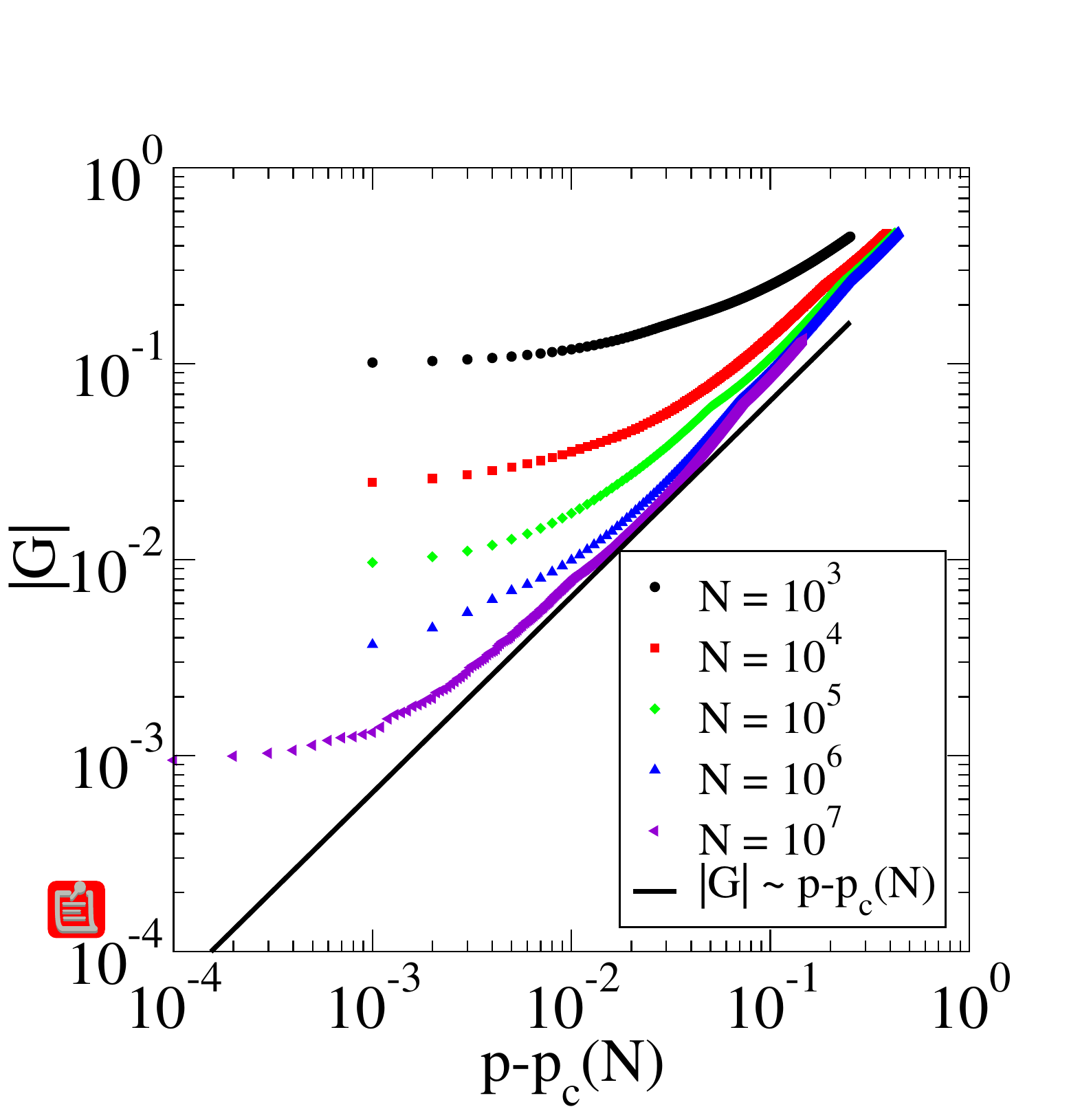}
  \caption{Relative size of the giant component $|G|$ vs $p-p_c(N)$
    for $\gamma=4.5$ and various $N$ compared with the analytical
    prediction.}
  \label{|G|4.5}
\end{figure}

In Figure~\ref{|G|4.5} we plot the relative size of the giant
component as a function of $p-p_c(N)$. Also here the analytical
prediction $|G| \sim \Delta$ works well but not perfectly. The effect
of the degree degeneracy between many nodes is witnessed by the
presence of little discontinuities in the slope of the curves,
corresponding to points where $k_p$ changes by a unit.

\subsection*{$3<\gamma<4$}

\begin{figure}
  \includegraphics[width=0.49\textwidth]{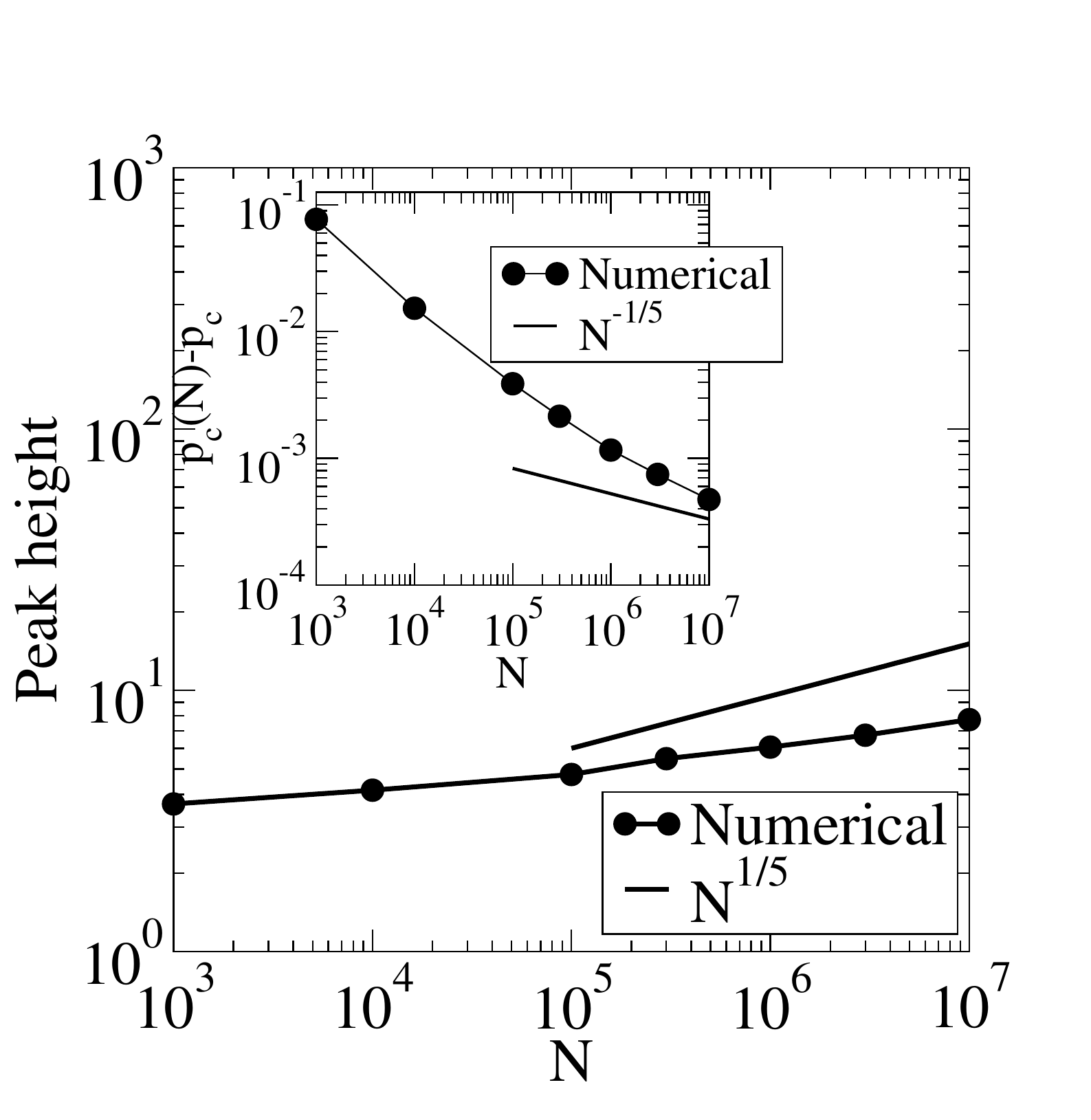}
  \caption{Results for $\gamma=3.5$.
    Main: Susceptibility peak height as a function of the system size
    $N$. Numerical results are compared with the theoretical prediction
    $N^{1-2\beta/\nu}$ which gives an exponent $1/5$
    in this range of $\gamma$ values.
    Inset: Difference between the numerical effective threshold $p_c(N)$ and the
  expected value for infinite size $p_c$ as a function of $N$. The straight solid line is the theoretical prediction $N^{-1/5}$.}
  \label{fss3.5}
\end{figure}

\begin{figure}
  \includegraphics[width=0.49\textwidth]{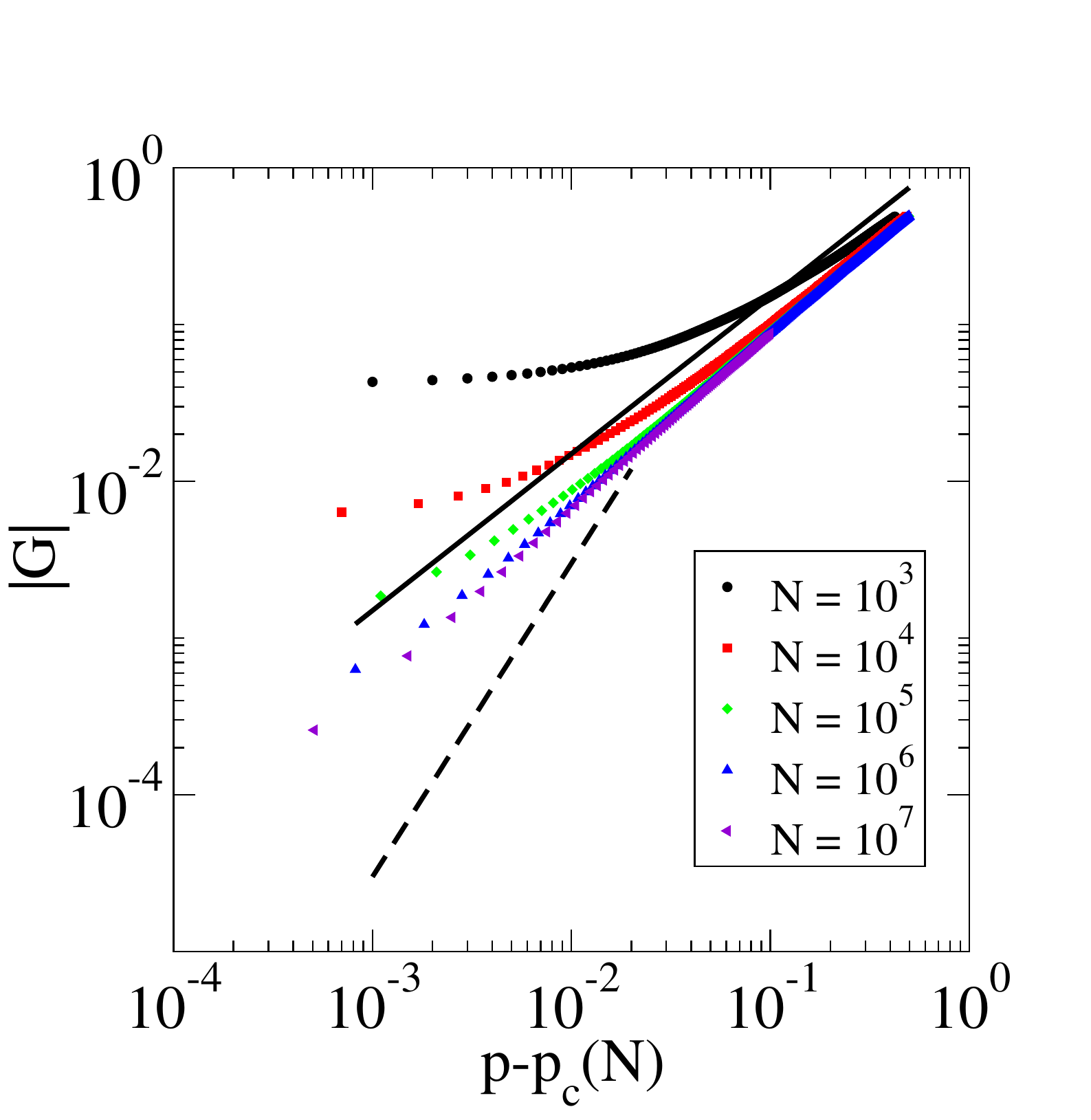}
  \caption{Relative size of the giant component $|G|$ vs $p-p_c(N)$
    for $\gamma=3.5$ and various $N$ compared with the analytical
    prediction, $|G| \sim \Delta^2$ (dashed line) and $|G| \sim
    \Delta$ (solid line).}
  \label{|G|3.5}
\end{figure}

\begin{figure}
  \includegraphics[width=0.49\textwidth]{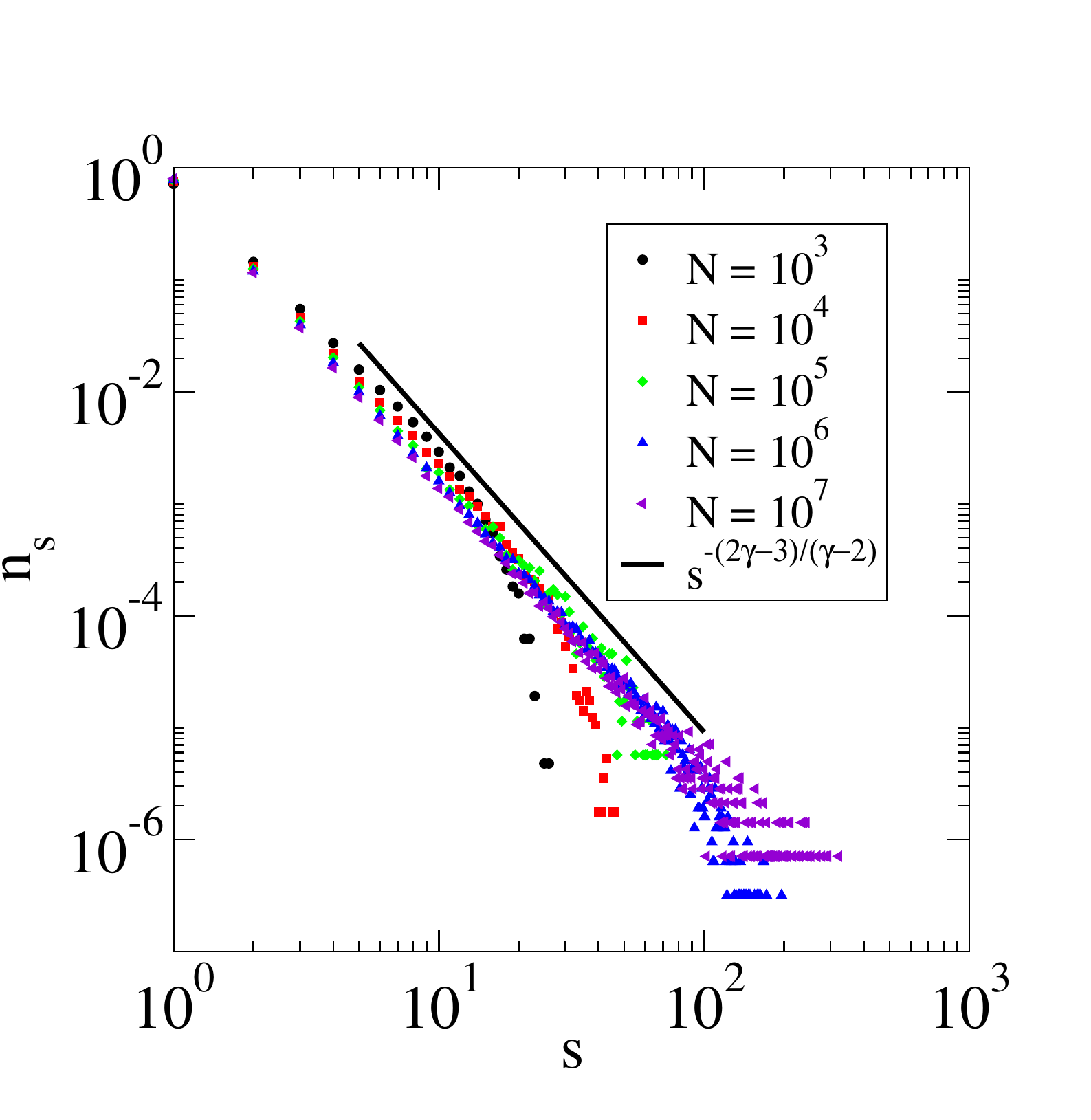}
  \caption{Cluster size distribution $n_s$ at the percolation
      threshold $p_c(N)$ for $\gamma=3.5$ and various system sizes $N$. The solid straight line
      represent the decay predicted
      analytically, Eq.~\eqref{tau}, which for $\gamma=3.5$ gives $\tau=8/3$.}
  \label{ns3.5}
\end{figure}

To verify the validity of analytical results in this range we consider
$\gamma=3.5$, for which the theoretical predictions are
$p_c=0.0000169$, $1/\nu=1/5$, $\beta=2$ and $\tau=8/3$.
Also for this value of $\gamma$ the threshold is finite. In principle,
because $k_c$ is larger, we should expect a better agreement between
the continuous theory and numerical results than in the case $\gamma=4.5$.
As shown below, however, this is not the case, because the threshold value
for infinite size is exceedingly small and the approach to it
is very slow, due to the large $\nu$ value. As a consequence huge finite
size corrections affect the results and in order to see the asymptotic
regime unfeasibly large values of $N$ would be needed.
A first evidence of this is provided by Fig.~\ref{fss3.5}, where
the scaling of the peak position (effective threshold) and of the
peak height with system size are displayed. The numerical curves
slowly approach the expected behavior, but much larger sizes would be
needed to see the truly asymptotic exponent.
A similar indication comes from the plot, in Figure~\ref{|G|3.5},
of the relative size of the giant component as a function
of $p-p_c(N)$. Even for the largest system size considered the effective
exponent is larger than 1 but definitely smaller than the expected value
$\beta=2$.
Instead the distribution of cluster sizes $n_s$ at the critical point
obeys well the expected behavior (see Fig.~\ref{ns3.5}).

\subsection*{$2<\gamma<3$}

\begin{figure}
  \includegraphics[width=0.49\textwidth]{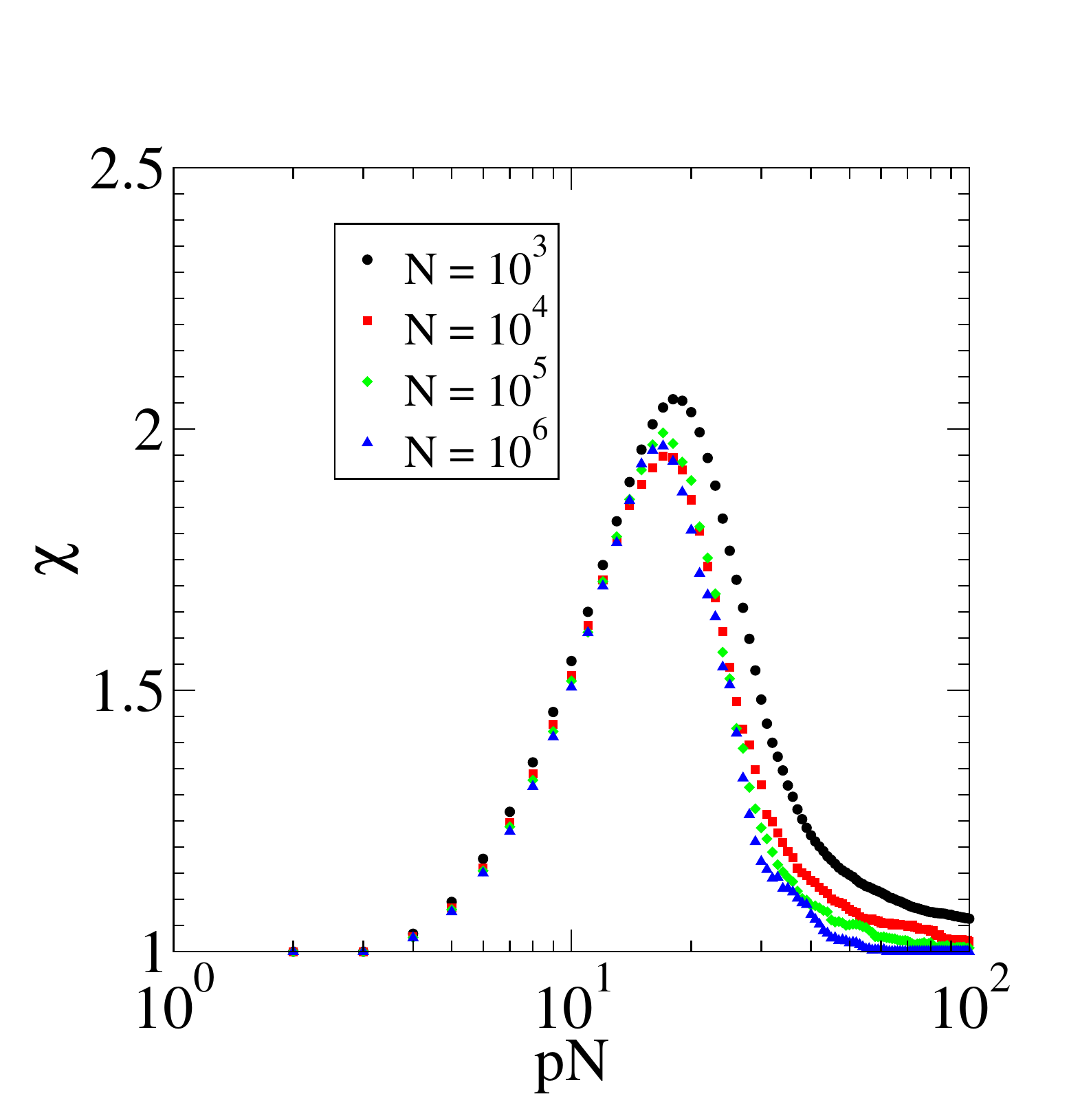}
  \caption{Plot of the susceptibility $\chi$ as a function of $pN$
    for $\gamma=2.5$ and various $N$.}
  \label{susc2.5}
\end{figure}

\begin{figure}
  \includegraphics[width=0.49\textwidth]{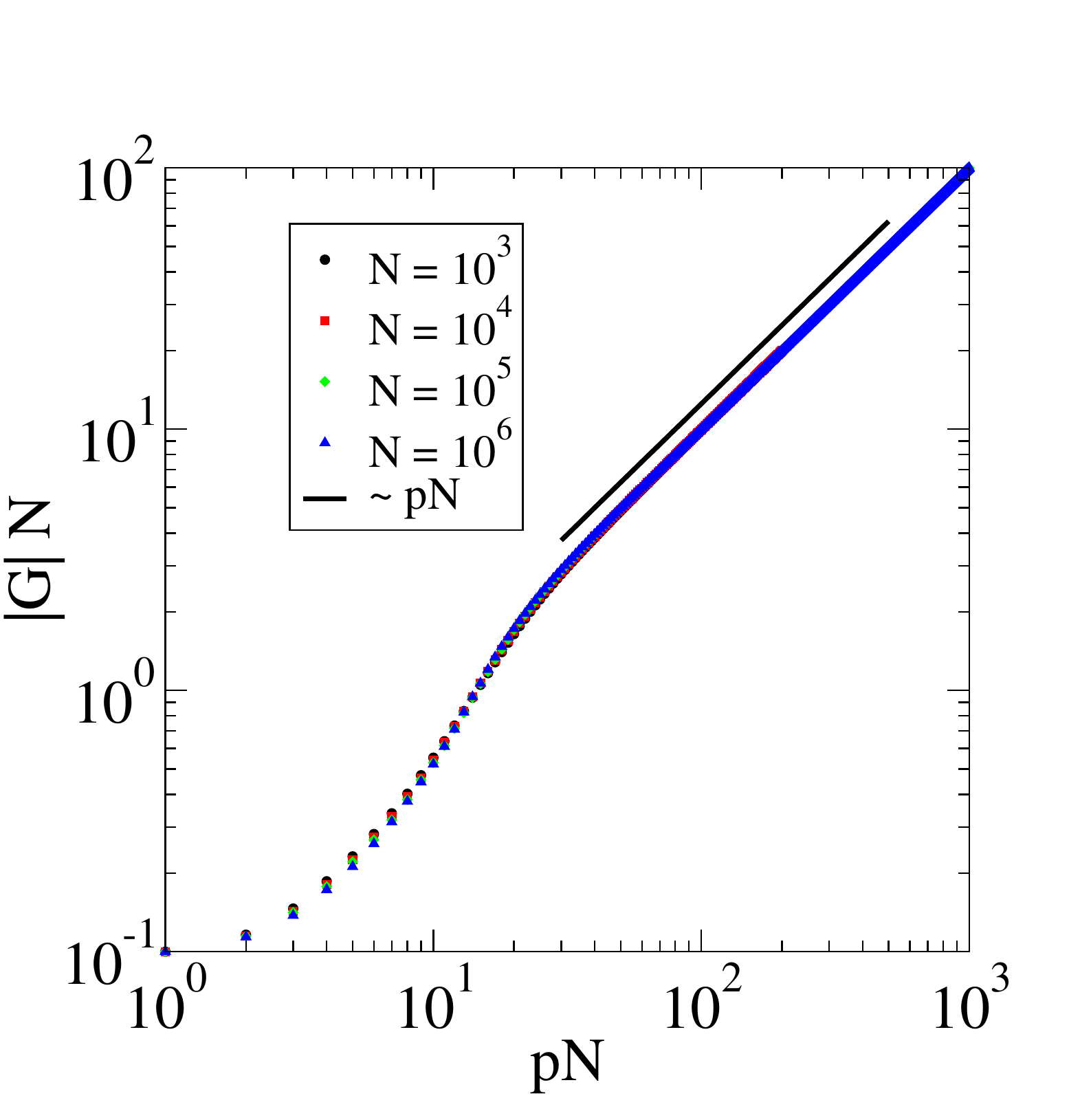}
  \caption{Plot of $|G|N$ vs $p$ for $\gamma=2.5$ and various $N$ compared
  with the analytical prediction.}
  \label{|G|2.5}
\end{figure}

For networks with $\gamma=2.5$ our theoretical approach predicts a
vanishing threshold in the infinite size limit and the
exponents $\nu=1$, $\beta=1$.
To test the validity of the prediction that $p_c(N) \sim N^{-1}$,
in Fig.~\ref{susc2.5} we plot the susceptibility $\chi$
versus $pN$ for various system sizes.
The perfect collapse confirms the validity of the
finite size scaling analysis.
Note that $\chi_{max}$ does not depend on $N$. This disagrees with the
prediction $\chi_{max} \sim N^{1-2\beta/\nu} = N^{-1}$, showing that
the hyperscaling relation does not hold in this case.
Note also that, as predicted, the effective
transition occurs when the number $N p_c(N)$ of nodes added to the system
is not only finite, but also very small, of the order of 10.
For this reason, when the incipient giant component starts to appear,
finite clusters -- if any -- are extremely tiny (of size 1 or 2)
\change{
  and no power-law decay is observed for their size.
}
Also the plot in Figure~\ref{|G|2.5}, displaying
the rescaled size $|G|N$ of the giant component as a function
of $pN$, shows a perfect agreement with the prediction $\beta=1$,
thus confirming the great accuracy of the theoretical predictions
for $2<\gamma<3$.

\section{Conclusions}

In summary, we have studied the transition of the Degree-Ordered-Percolation
model on power-law distributed uncorrelated networks. By applying
standard analytical methods we have determined the percolation threshold and
associated critical exponents as a function
of the exponent $\gamma$ of the degree distribution.
The results have then been checked by means of numerical simulations,
obtaining a satisfactory agreement except for the case $\gamma=3.5$ where
the discrepancy between theory and simulations can however be rationalized
as the effect of very strong finite size effects, associated to the
extremely small value of the threshold.
DOP is a variation of the standard random percolation process, which
exhibits nontrivial properties on heterogeneous networks.
A comparison of the results derived here with corresponding values
for standard percolation~\cite{Cohen2002} indicates that for
scale-rich topologies ($\gamma>3$) DOP is in the same universality
class, sharing the same critical exponents values.
It is however important to remark that the different protocols for removing
nodes have a strong influence on the value of the percolation thresholds,
which are very different in the two cases. As can be seen from the curves
in Fig.~\ref{pcvsgamma}, the threshold for standard percolation grows
(linearly) large as soon as $\gamma>3$, while DOP threshold remains
practically indistinguishable from $0$ for $\gamma$ up to 4.
This has important consequences for the SIS dynamics on this type
of networks, whose large-scale properties depend on the Cumulative
Merging Percolation process, which is a long-range variation of
DOP~\cite{Castellano2020}.

Our results confirm and clarify the numerical evidence presented by
Lee {\em et al.}~\cite{Lee2013}.
The singular behavior of the DOP threshold for $\gamma \to 3$ from above
is an anticipation of the nontrivial behavior observed for
$2<\gamma \le 3$.
In this range, the transition occurs for $p_c=0$, as
for standard percolation, but DOP is not in the same universality class,
having different exponents, independent of $\gamma$.
The value $\nu=1$, governing the approach to zero of
the size-dependent effective threshold $p_c(N)$, is quite peculiar.
It implies that a giant component starts forming as soon as a
fixed number (not a fixed fraction) of nodes are added.
Such a number turns out to be very small,
of the order of a few units, increasing further the oddity of this transition.

In the present paper we have investigated DOP on an ensemble of
random uncorrelated networks, where the only preassigned
property is the degree distribution. It is natural to wonder what is
the effect of additional topological features on this type of transition.
Among these possible further developments, a particularly interesting one
is the investigation of the effect of degree correlations. It is reasonable
to expect that assortative correlations will lower the threshold,
while disassortative ones will tend to increase it, since they will
make hubs more distant from each other.
Whether these tendencies lead to qualitative changes
(i.e., a vanishing threshold for $\gamma>3$ or
a finite threshold for scale-free networks)
is a nontrivial question that remains open.

\section*{Acknowledgments}
C. C. is grateful to Romualdo Pastor-Satorras for many discussions
on this topic.


%

\appendix

\section{Calculation of the exponent $\nu$ for $\gamma\, \change{\ge} \, 3$}
\label{fss}

For a network of finite size, the integrals in Eq.~\eqref{MR2} must
be performed only up to $\kM$, yielding \change{for $\gamma>3$}
\begin{widetext}
\be
\frac{A}{\av{k}}
\gra*{\frac{1}{3-\gamma} \qua*{\kM^{3-\gamma}-k_c(N)^{3-\gamma}}-
\frac{1}{2-\gamma} \qua*{\kM^{2-\gamma}-k_c(N)^{2-\gamma}}} = 1,
\ee
\end{widetext}
where $A$ is the normalization prefactor appearing in Eq.~\eqref{pk},
and we have noted explicity that now $k_c$ depends on $N$.
This equation can be rewritten as
\begin{widetext}
\be
\frac{\gamma-2}{\gamma-3} \km^{\gamma-2} k_c(N)^{2-\gamma}
\gra*{k_c(N) \qua*{1- \ton*{\frac{\kM}{k_c(N)}}^{3-\gamma}}-
\frac{\gamma-3}{\gamma-2}\qua*{1- \ton*{\frac{\kM}{k_c(N)}}^{2-\gamma}}
}= 1,
\label{kc2}
\ee
\end{widetext}
where we have already taken the large $N$ limit in the expression of $A$.
For $\kM \to \infty$, Eq.~\eqref{kc2} correctly returns Eq.~\eqref{kc}.
For finite $\kM$ the dominant correction is given by the term
$[\kM/k_c(N)]^{\gamma-3}$.
Inserting the assumption $k_c(N) = k_c + \delta$ into Eq.~\eqref{kc2}
(where $k_c$ is the solution of Eq.\eqref{kc}) and expanding for small
$\delta$ we find that $\delta \sim \kM^{3-\gamma}$.
At this point we can go back to Eq.~\eqref{pcf}, which is the equation
for $p_c$ for finite $\kM$.
Inserting $p_c(N)=p_c+\delta p_c$ into it we find
\be
\delta p_c \sim \delta \sim \kM^{3-\gamma} \sim N^{-(\gamma-3)/(\gamma-1)}.
\ee

\change{
  For $\gamma=3$ instead, Eq.~\eqref{MR2} gives
  \be
  \km \left[ \ln\left(\frac{\kM}{\km} \right) + \frac{1}{\kM} - \frac{1}{k_c}
    \right] = 1.
  \ee
  Neglecting the terms $1/\kM$ and $1/k_c$ which vanish in the thermodynamic limit,
  we find
  \be
  k_c = \kM e^{-\frac{1}{\km}}.
  \ee
  Inserting this expression into Eq.~\eqref{pc} we find
  \be
  p_c = \km^2 (k_c^{-2}-\kM^{-2}) = \km^2 \left(e^{2/\km}-1 \right) \kM^{-2}.
  \ee
  Since $\kM = N^{1/2}$ we finally obtain
  \be
  p_c = \km^2 \left(e^{2/\km}-1 \right) N^{-1}.
  \ee
  Hence we find that the exponent $1/\nu=1$ is the same of the case $2<\gamma<3$,
  while the prefactor has a different dependence on $\km$. In particular, there
  is a discontinuity between the  limit of the prefactor for $\gamma \to 3^-$,
  $2 \km$ and the value $\km^2 \left(e^{2/\km}-1 \right)$ for $\gamma=3$.

  It is also interesting to consider the limit $\gamma \to 3^+$
  {\em before} the limit $N \to \infty$.
  In that case one finds from Eq.~\eqref{MR2}
  \be
  k_c = \left(\frac{\gamma-3}{\gamma-2} \km^{2-\gamma}+\kM^{3-\gamma}
  \right)^{1/(3-\gamma)}
  \ee
  Inserting this into Eq.~\eqref{pc2} and taking the limit
  $\epsilon=\gamma-3 \to 0$ we find
  \be
  p_c = \left(\frac{\km}{\kM} \right)^2
  \left[1+ \frac{\epsilon}{\km} \left(\frac{\km^2}{\kM^2} \right)^{\frac{\epsilon}{2}}
    \right]^{\frac{2}{\epsilon}} \to \frac{\km^2}{\kM^2} e^{\frac{2}{\km}}.
  \ee
  Hence $p_c$ vanishes as $1/N$
  with a prefactor $k_{min}^2 e^{2/k_{min}}$ again discontinuous with respect
  to the case
  the case $\gamma=3$.
}

\section{Calculations for the exponent $\beta$}
\label{appbeta}
Let us consider Eq.~\eqref{G} and write it explicitly in the continuous degree
limit for a network of infinite size
\bea
|G| &=& f_0(1)-f_0(u) = p - \int_{k_p}^{\infty} dk P(k) u^k \\
& =& p - (\gamma-1)\km^{\gamma-1} \int_{k_p}^{\infty} dk k^{-\gamma} e^{k \ln u}.
\eea
Setting $t=k \ln(1/u)$ this can be rewritten as
\be
|G| = p - (\gamma-1)\km^{\gamma-1} [\ln (1/u)]^{\gamma-1}
\Gamma[1-\gamma,k_p \ln (1/u)],
\label{GGG}
\ee
where $\Gamma(a,z)$ is the incomplete Gamma function.
Setting $u=1-\epsilon$ and expanding for small $\epsilon$,
$\ln (1/u) = \epsilon + \epsilon^2/2$, we can use the expansion
of the Gamma function for $z \to 0$
\be
\Gamma(a,z)=\Gamma(a)-\frac{z^{a}}{a} \left[1-\frac{a z}{a+1}+\frac{a z^2}{2 (a+2)}+ \mathcal{O} \left( z^3 \right) \right].
\label{Gamma}
\ee
In this way we obtain
\bea \nonumber
|G(p)| &=& p- (\gamma-1)\km^{\gamma-1}\Gamma(1-\gamma)\epsilon^{\gamma-1}\\
&-&\ton*{\frac{\km}{k_p}}^{\gamma-1}
\ton*{1-\frac{1-\gamma}{2-\gamma} k_p \epsilon}.
\label{GG}
\eea
Since the degree $k_p$ is related to $p$ by
\be
p = \int_{k_p}^{\infty} dk P(k) = \ton*{\frac{\km}{k_p}}^{\gamma-1},
\ee
the first and the third term in Eq.~\eqref{GG} simplify. 
For any $\gamma>2$ the leading order in $\epsilon$ is then
\be
|G(p)| \simeq \frac{\gamma-1}{\gamma-2} \km^{\gamma-1} k_p^{2-\gamma} \epsilon.
\ee
The actual behavior of $|G|$ as a function of $\Delta=p-p_c$ depends
hence on how $k_p$ and $\epsilon$ depend in their turn on $\Delta$.

For $2<\gamma<3$, for which $p_c=0$, $\Delta=p$ and therefore
\be
k_p \sim \Delta^{-1/(\gamma-1)}.
\label{kp}
\ee

\change{
  For $\gamma=3$ the expansion of the function $\Gamma(-2,z)$
  in Eq.~\eqref{GGG} is different from Eq.~\eqref{Gamma},
  but this does not really lead to a modification of the result, which is
  $k_p \sim \Delta^{-1/2}$, i.e., Eq.~\eqref{kp} evaluated
  for $\gamma=3$.
  }

For $\gamma>3$ instead, since $p_c$ is finite, $k_p$ is also finite
(and equal to $k_c$) at the transition, so that close to it we can write
\be
k_p = k_c + a\Delta,
\label{kpp}
\ee
with $a = \km p_c^{\gamma/(1-\gamma)}/(1-\gamma)$.

Concerning $\epsilon$ instead, Eq.~\eqref{u} for $u$ reads
\be
u = 1+\int_{k_p}^\infty dk \frac{k P(k)}{\av{k}} (u^{k-1}-1),
\ee
that can be rewritten as
\bea
u &=& 1-\km^{\gamma-2}k_p^{2-\gamma}+ \\ \nonumber
&+&(\gamma-2)\km^{\gamma-2} \frac{1}{u}
\qua*{\ln\ton*{1/u}}^{\gamma-2} \Gamma(2-\gamma,k_p \ln\ton*{1/u}).
\label{uu}
\eea

Setting $u=1-\epsilon$, expanding the incomplete Gamma function for
small values of the second argument, using $1/u \approx 1+\epsilon$
and $\ln(1/u) \approx \epsilon + \epsilon^2/2$ and keeping only lowest
order terms we finally arrive at
\bea \nonumber
\frac{\epsilon}{\km^{\gamma-2}} &=& -(\gamma-2) \Gamma(2-\gamma) \epsilon^{\gamma-2}
- k_p^{2-\gamma} \epsilon + \frac{2-\gamma}{3-\gamma} k_p^{3-\gamma} \epsilon \\
&+&
\qua*{\frac{3}{2} \frac{(2-\gamma)}{(3-\gamma)} k_p^{3-\gamma} -
  \frac{2-\gamma}{2(4-\gamma)} k_p^{4-\gamma}}\epsilon^2.
\label{ss}
\eea

For $2<\gamma<3$ the leading terms are $\epsilon^{\gamma-2}$ and
$\epsilon k_p^{3-\gamma}$. Imposing that they balance each other
asymptotically implies
\be
\epsilon \sim k_p^{-1}.
\ee

For $\gamma>3$, inserting into Eq.~\eqref{ss} the expansion~\eqref{kpp}
of $k_p$ and using the threshold condition~\eqref{kc} we obtain
\bea \nonumber
0 &=& -  (\gamma-2) \Gamma(2-\gamma) \epsilon^{\gamma-2}-a (2-\gamma)
k_c^{1-\gamma}\epsilon\Delta + \\
&+& \frac{2-\gamma}{3-\gamma} \qua*{
  (3-\gamma)a k_c^{2-\gamma} \epsilon \Delta + \frac{3}{2} k_c^{3-\gamma} \epsilon^2}
+ \\ \nonumber
&-& \frac{2-\gamma}{2(4-\gamma)} k_c^{4-\gamma} \epsilon^2.
\eea
If $\gamma>4$ the leading terms on the r.h.s. are those proportional to
$\epsilon \Delta$ and $\epsilon^2$. Their matching implies
\be
\epsilon \sim \Delta.
\ee
If $3<\gamma<4$ the leading terms are $\epsilon^{\gamma-2}$ and $\epsilon \Delta$,
implying
\be
\epsilon \sim \Delta^{1/(\gamma-3)}.
\ee

\change{
Finally, for $\gamma=3$ the expansion of the $\Gamma$ function in Eq.~\eqref{uu}
is
\be
\Gamma(-1,z)=\ln(z)+ \frac{1}{z}+ \left( \gamma_E -1 \right)+ \mathcal{O} \left( z \right).
\ee
Expanding for $\epsilon \to 0$ and neglecting higher order terms we obtain
\be
\km \ln(k_p \epsilon) = \km(1-\gamma_E-1/k_p)-1
\ee
When $\epsilon \to 0$, for the l.h.s. to be finite it must be $\epsilon \sim k_p^{-1}$,
as in the case $2<\gamma <3$. As a consequence the behavior of the giant
component is characterized by the same value $\beta=1$.
}

\section{Calculations for the exponent $\tau$}
\label{C}

Eq.~\eqref{h1} reads \change{for $\gamma>3$}
\be
h_1 = 1-\int_{k_c}^{\infty} dk P(k) \frac{k}{\av{k}} + (1-\epsilon)
\int_{k_c}^{\infty} dk P(k)  \frac{k}{\av{k}} h_1^{k-1}.
\label{C1}
\ee

Setting $t=k\ln(1/h_1)$ we obtain
\bea \nonumber
h_1 & = & 1- \km^{\gamma-2} k_c^{2-\gamma}
+(1-\epsilon) (\gamma-2) \km^{\gamma-2}\frac{1}{h_1} \cdot \\
& \cdot & [\ln(1/h_1)]^{\gamma-2} \Gamma[2-\gamma,k_c \ln(1/h_1)].
\eea

Close to the transition, we take $x=1-\epsilon$ and
define the function $g_1(\epsilon)=1-h_1(1-\epsilon)$.
$g_1$ is small and we can expand $1/h_1 \approx 1 + g_1 + \ldots$
and $\ln(1/h_1) \approx g_1 + g_1^2/2 + \ldots$.
Furthermore the incomplete Gamma function can be expanded
for small values of the second argument, as in Eq.~\eqref{Gamma}.
After straightforward algebra, by using the condition~\eqref{kc} 
and neglecting all subleading terms, we obtain
\bea
\epsilon k_c^{2-\gamma} &=& (\gamma-2)\Gamma(2-\gamma) g_1^{\gamma-2} \\ \nonumber
&+& \qua*{k_c^{\gamma-2}- \frac{3}{2} \frac{\gamma-2}{\gamma-3} k_c^{3-\gamma}+
\frac{2-\gamma}{2(4-\gamma)}k_c^{4-\gamma}} g_1^2.
\eea
For $\gamma>4$ the leading term on the r.h.s. is the one proportional
to $g_1^2$,
implying that $g_1 \sim \epsilon^{1/2}$.
For $\gamma<4$ the leading term is the one proportional to $g_1^{\gamma-2}$,
so that $g_1 \sim \epsilon^{1/(\gamma-2)}$.
\change{
  For $2 < \gamma \le 3$ performing the integrals in Eq.~\eqref{C1} between
  $k_c$ and $k_{max}$ and letting $N \to \infty$ both integrals vanish.
  Trivially $h_1=1$ for any $\epsilon$. The probability
  that a finite cluster has size $s$ does not decay as a power-law.
}

In order to determine the exponent $\tau$ we must determine the critical properties of the generating function $h_0(x)$.
Defining $g_0(\epsilon) = 1 - h_0(1-\epsilon)$
and inserting it into Eq.~\eqref{h0} we obtain
\bea
g_0 & = & \km^{\gamma-1} k_c^{1-\gamma} - (1-\epsilon)(\gamma-1) \km^{\gamma-1} \cdot \\
 \nonumber
& \cdot & \ln[(1-g_1)^{-1}]^{\gamma-1} \Gamma\ton*{1-\gamma,k_c \ln[(1-g_1)^{-1}]}.
\eea
By expanding the incomplete Gamma function, reminding the expansions of
the terms containing $1-g_1$ and neglecting subleading terms, we arrive at
\bea \nonumber
\frac{g_0}{\km^{\gamma-1}} &=& -(1-\epsilon)(\gamma-1)\Gamma(1-\gamma) g_1^{\gamma-1} +
\epsilon k_c^{1-\gamma} \\
&-& (1-\epsilon) \frac{\gamma-1}{2-\gamma}k_c^{2-\gamma} g_1.
\label{g0}
\eea
For $\gamma>4$, since $g_1 \sim \epsilon^{1/2}$ the leading term is the third
and we have
\be
g_0 \sim g_1 \sim \epsilon^{1/2}.
\ee
For $3<\gamma<4$ instead $g_1 \sim \epsilon^{1/(\gamma-2)}$. The leading term
is still the third, resulting in
\be
g_0 \sim g_1 \sim \epsilon^{1/(\gamma-2)}.
\ee
\end{document}